\documentstyle[12pt]{article}
\begin{document}

\begin{titlepage}
\begin{center}
{\Large Analysis of Charged-Particle/Photon\\ Correlations in Hadronic
Multiparticle
Production}
\end{center}

\begin{center}
T. C. Brooks, M. E. Convery, W. L. Davis, K. W. DelSignore,$^{1}$ T. L.
Jenkins,\\
 E. Kangas,$^{2}$ M. G. Knepley,$^{3}$  K. L. Kowalski, and C.C. Taylor\\  
{\footnotesize Department of Physics, Case Western Reserve University,
 Cleveland, Ohio 44106}
\end{center}

\begin{center}
S. H. Oh and W.D. Walker\\
{\footnotesize Department of Physics, Duke University, Durham, North
Carolina 27708-0305}\\
\end{center}

\begin{center}
P. L. Colestock, B. Hanna, M. Martens, and J. Streets\\
{\footnotesize Fermi National Accelerator Laboratory, P.O. Box 500,
 Batavia, Illinois 60510}
\end{center}

\begin{center}
R. Ball, H. R. Gustafson, L.W. Jones, and M. J. Longo\\
{\footnotesize Department of Physics, University of Michigan, 
Ann Arbor, Michigan 48109-1120}
\end{center}

\begin{center}
J. D. Bjorken\\
{\footnotesize  Stanford Linear Accelerator Center, Stanford University, 
Stanford, California  94305}
\end{center}

\begin{center}
A. Abashian and N. Morgan\\
{\footnotesize  Department of Physics, Virginia Polytechnic Institute,
Blacksburg, 
Virginia 24061-0435}
\end{center}

\begin{center}
C. A. Pruneau\\
{\footnotesize Department of Physics and Astronomy, Wayne State University, 
Detroit, Michigan 48202}
\end{center}

\begin{center}
(MiniMax Collaboration)
\end{center}

\begin{abstract}
In order to analyze data on joint charged-particle/photon distributions
from an experimental
 search (T-864, MiniMax) for disoriented chiral condensate (DCC) at the
Fermilab Tevatron
 collider, we have identified robust observables, ratios of normalized
bivariate factorial
 moments, with many desirable properties. These include insensitivity to
many efficiency
 corrections and the details of the modeling of the primary pion
production, and sensitivity
 to the production of DCC, as opposed to the generic, binomial-distribution
partition of
 pions into charged and neutral species. The relevant formalism is
developed and tested in
 Monte-Carlo simulations of the MiniMax experimental conditions.
\\
\\
\\
\\
\\
\\
\\
\\
\\
\end{abstract}
{\footnotesize $^{1}$Now at Department of Physics, University of Michigan, 
Ann Arbor, MI 48109-1120}\\
{\footnotesize $^{2}$Now at Department of Physics, Massachusetts Institute of
Technology, Cambridge, MA 02139}\\
{\footnotesize $^{3}$Now at Department of Computer Science, University of
Minnesota, 
Minneapolis, MN 55455}

\end{titlepage}

\section{INTRODUCTION}
There has recently been renewed interest in semiclassical mechanisms of
pion production in 
high-energy collisions of hadrons and of heavy ions [1-11]. One hypothesis
in particular is 
that pieces of strong-interaction vacuum with 
an unconventional orientation of the chiral order parameter may be produced
in high energy 
collisions [12]. This disoriented chiral 
condensate (DCC) is then supposed to decay into a coherent semiclassical
pion field having the
 same chiral orientation.

The primary signature of this mechanism is the presence of large,
 event-by-event fluctuations in the fraction, 
$f$, of produced pions that are neutral. Conventional mechanisms of
particle production, 
including those used in standard Monte Carlo simulations, predict that the
partition of pions 
into charged and neutral species is governed by a binomial distribution
which, in the limit of 
large multiplicity, leads to a sharp value of $f \approx 1/3$. We refer to
this as 
{\em generic} pion production. On the other hand, for the decay of a pure
DCC state the
 distribution of neutral fraction is very different, following an inverse
square-root law 
in the limit of large multiplicity [1-7, 12]. Some other production
scenarios involving the 
common feature of coherent final states lead to identical $f$ distributions
[9, 10, 13-15].

Sophisticated phenomenological techniques have been developed in order to
study the properties of
 multiparticle final states, and much has been done on multiplicity distributions, 
correlations, and fluctuations  [16-20]. Most of the practical studies,
however, have considered 
the properties of a single species at a time. In the
 case of DCC, formal tools for the study of the joint distribution of
neutral and charged pions are
 required, and here there is much less data and corresponding analysis
experience [21-25]. 

The authors of this paper comprise the MiniMax collaboration (Fermilab T-864),
 who for the last three years
 have carried out an exploratory search for signals of DCC at the C0 area
of the Tevatron 
collider [26]. The heart of our detector is a telescope of 24 multiwire
 proportional chambers (MWPC),
 with a 1.0 radiation-length lead converter inserted after the eighth
 plane, so that charged tracks and converted photons can be counted
event-by-event. 
The acceptance in the lego space of pseudorapidity, $\eta$, and azimuthal
angle, $\phi$,
 is roughly a circle of radius 0.65 centered at $\eta = 4.1$. In 1995-1996,
8 million
 triggered events at $\sqrt{s} = 1.8$ TeV were recorded. The purpose of
this paper is not 
to report the results of this experiment, but rather to describe the
techniques we are using
 as the basis of our data analysis strategy. We believe these techniques
have much wider 
applicability and may be of value in other searches for DCC signals.

Even from this very brief description of the experiment, it should be clear
that we face many 
challenges in trying to infer either the presence or absence, within
limits, of DCC signals 
from the data. These include the following:
\begin{itemize}
\item[(a)] The MiniMax acceptance is small, so that it is improbable that 
both $\gamma$'s from a $\pi^{0}$ enter the detector acceptance.
\item[(b)] The conversion efficiency per $\gamma$ is about 50\%. 
\item[(c)] Not all $\gamma$'s come from $\pi^{0}$'s.
\item[(d)] Not all charged tracks come from $\pi^{\pm}$'s. 
\item[(e)] Because of the small acceptance, the multiplicities are rather low, 
so that statistical fluctuations are very important. 
\item[(f)] Detection efficiencies for charged tracks and $\gamma$'s are 
momentum-dependent and are not the same.
\item[(g)] Efficiency functions may be dependent upon the observed
multiplicity or 
other parameters.
\item[(h)] The efficiency for triggering when no charged track 
or converted $\gamma$ is produced within our acceptance is relatively low
and different
from that for events in which at least one charged particle or converted
$\gamma$ is detected. 
\end{itemize}

Nevertheless, we find that there do exist observables which are robust in
the sense that,
 even in the presence of large (uncorrelated) efficiency corrections and
convolutions from
 produced $\pi^{0}$'s to observed $\gamma$'s, the observables take very
different values
 for pure DCC and for generic particle production. Each such observable is
a ratio,
 collectively referred to as $R$, of certain bivariate normalized factorial
moments,
 that has many desirable properties, including the following:

\begin{enumerate}
\item The $R$'s do not depend upon the form of the parent pion multiplicity 
distribution.
\item The $R$'s are independent of the detection efficiencies for finding
charged 
tracks, provided these efficiencies are not correlated with each other or
with other 
variables such as total multiplicity or background level.
\item Some of the $R$'s are also independent of the $\gamma$ efficiencies 
in the same sense as above. In the remaining cases, the $R$'s depend only
upon one parameter,
 $\xi$, which reflects the relative probability of both photons from a
$\pi^{0}$ being 
detected in
 the same event.
\item In 
all cases $R$ is independent of the magnitude of the null trigger
efficiency; see comment (h) above. 
\item The ratios $R$ possess definite and very different values for pure
generic and 
pure DCC pion production.
\end{enumerate}

The idealizations implicit in the realization of properties 1-5 include the
assumptions that
 particles other than pions can be ignored, that there is no
misidentification of charged 
particles with photons, and that the production process can be modeled as a
two-step process,
 with a parent-pion multiplicity distribution posited, followed by a particular 
charged/neutral partitioning of that population by, e.g., a binomial or DCC
distribution 
function. In addition, there is the vital assumption that detection
efficiencies for finding
 a $\pi^{\pm}$ or $\gamma$ do not depend upon the nature of the rest of the
event. The validity
 of these idealizations is not contradicted by the simulations presented in
this paper. This 
idealized model thus appears to be a good basis for a first-order analysis
of the properties of 
the ratios $R$. We anticipate that this will remain true for observations
more general
 than those of the MiniMax experiment. 

The layout of this paper is as follows: In Section 2 we review the
conventional formalism [16-20] of
 single-variable generating functions and factorial moments used in
describing global 
multiplicity distributions. We then develop the extensions required to
describe the bivariate
 case of distributions of $\pi^{\pm}$'s and $\pi^{0}$'s. The 
modifications needed to accommodate the decay of $\pi^{0}$'s into
$\gamma$'s, as well as the
 inclusion of less-than-perfect detection efficiencies for charged tracks
and $\gamma$'s, are 
considered in Section 3. 
In Section 4 we introduce the robust observables $R$ and demonstrate their
sensitivity to
 charged-particle/photon correlations and their insensitivity to detection 
inefficiencies and the overall aspects of the primary production process
for a wide class
 of production models. The DCC distribution is shown to fall into that
class, but with 
distinctly different values of the $R$'s that clearly distinguish it from
the generic 
distribution under realistic experimental conditions. Generalizations of
the formalism 
which allow for the admixture of both generic and DCC charged/neutral
production are 
considered in Section 5. In Section 6 we 
estimate, by Monte Carlo simulation as well as  by use of the UA5
charged-particle/photon data at 
200 GeV and 900 GeV [25], the effects on the $R$'s from the realistic 
complications discussed in the preceding paragraph. Concluding remarks are
made in 
Section 7.  A number of new results concerning the interpretation and 
representation of the standard DCC probability distribution that are needed
to establish our 
results concerning DCC production are presented in the Appendix.

\section{GENERATING FUNCTIONS FOR CHARGED-PION/NEUTRAL-PION DISTRIBUTIONS}
The entire content of a set of probabilities $\{P(N)\}$ for the production
of $N$ particles in a 
fixed region of phase space can be encapsulated into the generating function
\begin{equation}
G(z) =\sum\limits_{N=0}^{\infty}z^{N}P(N)
\end{equation}
whose derivatives evaluated at $z=1$ yield the factorial moments
\begin{eqnarray}
f_{i} &\equiv& \left({{d^{i}G(z)} \over {dz^{i} }}\right)_{z=1} \nonumber \\
      & = & \langle N(N-1) \cdots (N -i+1) \rangle.
\end{eqnarray}

It is often useful to express $P(N)$ as a Poisson transform [27] where one
introduces
 a spectral representation in terms of Poisson distributions with a
weighting function $\rho(\mu)$:
 \begin{equation}
P(N) = \int_{0}^{\infty} d\mu~\rho(\mu)~\frac{\mu^{N}}{N!}e^{-\mu},
 \end{equation} 
where  
\begin{equation}
\int_{0}^{\infty} d\mu~\rho(\mu)=1.
 \end{equation} 
The Poisson transform isolates the random statistical fluctuations from the
physics 
contained in $\rho(\mu)$. As an example, the negative binomial parametrization
\begin{equation}
\rho(\mu) ={{\lambda^{k}}\over{\Gamma(k)}} \mu^{k-1}~e^{-\lambda\mu},
\end{equation}
where $\lambda = k/\langle N \rangle$, gives a fairly good two-parameter
description of charged 
multiplicity distributions [16, 17]. From (1) and (3) we also obtain a
spectral representation
for the generating
function:
\begin{equation}
G(z) =  \int_{0}^{\infty}d\mu \, \rho(\mu) \, e^{\mu(z-1)},
\end{equation}
where now the factor $e^{\mu(z-1)}$ reflects the purely random character of
the Poisson distribution.

The generating function formalism has been widely used to study 
charged-hadron multiplicity distributions [16-20, 27]. We next generalize this 
formalism to bivariate distributions of charged and neutral pions. Among
our motivations for 
doing this is the simple manner in which 
detection inefficiencies and particle decays can be handled with generating
functions [27]. 
These features are particularly
important in dealing with the MiniMax experimental situation. 
Here the parent $\pi^{0}$'s are not reconstructed from the observed $\gamma$'s
 and the efficiencies for detecting
both the charged particles and the photons are less than perfect. 
These extensions are taken up in detail in
succeeding sections. Some earlier work in this connection is contained in
Refs. [21-25]

Let $p(n_{ch}, n_{0})$ denote the probability distribution  for the
occurrence of $n_{ch}$ and 
$n_{0}$ charged and neutral pions, respectively, in a multiparticle event
within a given 
phase-space region. As in the single-variable case, the content of this
bivariate distribution can be conveniently represented 
by the generating function for factorial moments defined by 
\begin{equation}   
G(z_{ch}, z_{0}) = \sum\limits_{n_{ch}, n_{0}=0}^{\infty} p(n_{ch}, n_{0})
z_{ch}^{n_{ch}} z_{0}^{n_{0}}. 
\end{equation} 
The partial derivatives of $G(z_{ch}, z_{0})$ evaluated at $z_{ch}=z_{0}=1$
generate the
 factorial moments referring to charged (ch) and neutral (0) particles:
\begin{equation}
f_{i,j}(ch, 0) \equiv \left({{\partial^{i,j}G(z_{ch}, z_{0})} \over
{\partial z_{ch}\,^{i} \partial z_{0}\,^{j}}}\right)_{z_{ch}= z_{0}=1}. 
\end{equation}
For example, we have, 
\begin{eqnarray}
f_{1,0}(ch, 0) & = & \langle n_{ch}\rangle, \nonumber \\
f_{0,1}(ch, 0) & = & \langle n_{0}\rangle, \nonumber \\
f_{1,1}(ch, 0) & = &  \langle n_{ch}n_{0}\rangle, \nonumber \\
f_{2,0}(ch, 0) & = & \langle n_{ch}(n_{ch}-1)\rangle.
\end{eqnarray}

Next, let $P(N)$ be the probability for producing a total of $N$ pions with
any distribution of 
charge among them. Then $p(n_{ch}, n_{0})$ can be written as the product of 
two disjoint probability distributions: 
\begin{equation}
            p(n_{ch}, n_{0}) =  P(N)\hat{p}(n_{ch}, n_{0}; N), 
\end{equation}
where $N = n_{ch} + n_{0}$, and 
\begin{equation}            
\sum\limits_{N=0}^{\infty}P(N)=1, 
\end{equation} 
\begin{equation}           
\sum\limits_{n_{ch}=0, n_{0}=0}^{\infty} \delta_{N , n_{ch} +
n_{0}}\hat{p}(n_{ch}, n_{0}; N) =  1. 
\end{equation}

What we call the generic
 model for the charged-neutral distribution $\hat{p}(n_{ch}, n_{0}; N)$ involves
 no correlations, namely, a binomial ({\em Bin}) 
distribution of $n_{ch}$ and $n_{0}$:
\begin{equation}
\hat{p}_{Bin}(n_{ch}, n_{0}; N) =\left( \begin{array}{c} 
N\\
n_{0}
\end{array}\right)
\hat{f}^{n_{0}}(1-\hat{f})^{n_{ch}}.
\end{equation}
Here $\hat{f}$ is the mean fraction  of $\pi^{0}$'s, which is expected to
be about $1/3$ as 
a consequence of isospin symmetry. If we substitute (13) into (10) and
explicitly denote the dependence on $\hat{f}$,
the generating function (7) becomes, in the binomial case, 
\begin{equation}   
G_{Bin}(z_{ch}, z_{0};\hat{f}) = \sum\limits_{N} P(N) [\hat{f}z_{0} +
(1-\hat{f}) z_{ch}]^{N}, 
\end{equation} 
which only depends on the linear combination
\begin{equation} 
\zeta \equiv \hat{f}z_{0} + (1-\hat{f}) z_{ch}.
\end{equation} 
 Conversely, if a generating function $G(z_{ch}, z_{0})$ 
is a function only of $\zeta$, the charged and 
neutral pions are binomially distributed.

If $P(N)$ is a Poisson 
distribution, $\ln G_{Bin}(z_{ch}, z_{0}; \hat{f})$ is linear in $\zeta$.
The simulations
 of generic production described in Section 6 yield 
generating functions that, to good approximation, depend only on a fixed 
linear combination of $z_{ch}$ and $z_{0}$; the incorporation of the
modeling of the MiniMax 
detector into these simulations is found to alter this linear behavior slightly.

Much of the simplicity of the generic case is also realized for
 what can be called the binomial transform 
\begin{equation}   
\hat{p}(n_{ch}, n_{0}; N) = \left( \begin{array}{c}
                                        N\\
                                        n_{0}
                                    \end{array}\right) \int_0^1
{{df}}p(f)f^{n_{0}}(1-f)^{n_{ch}}, 
\end{equation} 
of the normalized distribution $p(f)$, 
\begin{equation}  
 \int_0^1 {{df}}p(f) =1. 
\end{equation} 
This leads to a wide class of possible pion factorial-moment generating
functions, namely
\begin{equation}   
G(z_{ch}, z_{0}) =  \int_0^1 {{df}}p(f) G_{Bin}(z_{ch}, z_{0};f), 
\end{equation}
where $G_{Bin}(z_{ch}, z_{0};f)$ is given by (14) with $\hat{f}$ replaced
by an arbitrary $f$, $0 \leq f \leq 1$. 
Combining (3) and (14) we obtain
\begin{equation}
G(z_{ch}, z_{0}) =  \int_{0}^{\infty}d\mu \, \rho(\mu)\int_0^1 {{df}} \,
p(f) \,e^{\mu[\zeta (f)-1]},
\end{equation}
where again $\zeta (f) $ is given by (15) with $\hat{f}$ replaced by an
arbitrary $f$.

The forms of $p(f)$ and $\rho(\mu)$ depend on the production model and the
detector. The uncorrelated, 
generic case (14) corresponds to $p(f) = \delta(f-\hat{f})$, where
$\hat{f}$ is some 
fixed value of $f$. 

It is shown in the Appendix that for a simple DCC model [1-7] and with a
sampling 
prescription appropriate to the experimental situation, $p(f)=1/(2\sqrt
{f})$. Although the same
 bivariate distribution is realized in other hadronic production models 
leading to coherent states [9, 10, 13-15], we refer to this case as the DCC
model. We  note that in 
the DCC model $\langle n_{0} \rangle = 2\langle n_{ch} \rangle$, just as in
the generic case for 
$\hat{f} = 1/3$.

It is quite possible that the parent pion distribution $P(N)$ or,
equivalently, $\rho(\mu)$,
 will be
different for the DCC and generic production mechanisms. This distinction
is important 
for our considerations of admixtures of the two mechanisms. We investigate
some possible 
scenarios for such admixtures in Section 5.

\section{GENERATING FUNCTIONS FOR CHARGED-PION/PHOTON DISTRIBUTIONS }
For a detector that is designed to observe charged particles and converted
$\gamma$'s
 within its acceptance, events are classified only according to the numbers 
of charged particles and photons, $n_{ch}$ and $n_{\gamma}$, respectively.
With sufficiently
large statistics we can determine probabilities, $p(n_{ch},n_{\gamma})$,
for observing these 
combinations over some portion or all of the available phase space.

In order to obtain the charged-pion/photon generating function, 
incorporating both $\pi^{\pm}$ and $\gamma$ detection efficiencies   
from $G(z_{ch}, z_{0})$, we extend Pumplin's cluster 
theorem [27] to the bivariate case. Consider a generating function 
$G(z_{ch}, z_{0})$ that
 refers to charged and neutral ``clusters.'' Suppose, for the sake of
simplicity, the charged clusters 
decay in a number of
ways into charged particles and likewise for the decay of neutral clusters
into neutral 
particles. For each of these decay scenarios there is a probability
distribution and a corresponding 
generating function, $g_{ch}(z_{ch})$ or $g_{0}(z_{0})$, respectively. The
bivariate 
generating function of the factorial moments of the final charged-neutral
particle production is then
$G(g_{ch}(z_{ch}), g_{0}(z_{0}))$. If the charged clusters do not decay, then 
$g_{ch}(z_{ch})=z_{ch}$. On the other hand, $\pi^{0} \rightarrow \gamma \gamma$
with perfect photon detection efficiency correponds to 
$g_{0}(z_{\gamma})=z^{2}_{\gamma}$.    

More realistically, there is a probability $\epsilon_{ch}$ for observing a
given primary 
charged pion in the detector and a probability $1-\epsilon_{ch}$ for not
observing it. These possibilities
can be regarded as the two ``decay'' modes of the primary charged pion
which is otherwise regarded
as stable. Similarly,
 there are probabilities $\epsilon_{m}$, $m =0, 1, 2$, with
\begin{equation}
\epsilon_{0} + \epsilon_{1} + \epsilon_{2} = 1,
\end{equation} for observing $m$ 
photons from a $\pi^{0}$ decay and each possibility
can be regarded a decay mode of the $\pi^{0}$ cluster. If these probabilities 
are identified with what we suppose are the independent, i.e., uncorrelated, 
efficiencies for the respective detection options, the 
generating function for the distribution of observed particles, including
efficiencies,
is obtained from 
$G(z_{ch}, z_{0})$ by replacing $z_{ch}$ by the generating function
\begin{equation}
g_{ch}(z_{ch})= (1-\epsilon_{ch}) + \epsilon_{ch} z_{ch} ,
\end{equation} 
 and $z_{0}$ by the generating function
\begin{equation}
g_{0}(z_{\gamma})=\epsilon_{0} +\epsilon_{1} z_{\gamma} +\epsilon_{2}
z_{\gamma}^{2}.
\end{equation}

For the class of production models characterized by (18), the preceding
considerations 
lead
to the following factorial-moment generating function for the distribution
of observed charged pions and photons:
\begin{equation}
  G_{obs}(z_{ch}, z_{\gamma}) = \int_0^1 {{df}}p(f)G_{Bin}(g_{ch}(z_{ch}),
g_{0}(z_{\gamma});f).
\end{equation}
The charged-pion/photon factorial moments are
\begin{equation}
f_{i,j}(ch, \gamma) \equiv \left({{\partial^{i,j}G(z_{ch}, z_{\gamma})}
\over {\partial z_{ch}\,^{i} \partial z_{\gamma}\,^{j}}}\right)_{z_{ch}=
z_{\gamma}=1}, 
\end{equation}
which introduces the bivariate indexing $(i,j)$ with respect to charged
particles and photons  
employed henceforth.
For example, the two lowest orders of factorial
moments are:
\begin{equation}
f_{1,0}(ch, \gamma) =\langle n_{ch}\rangle =\langle1-f\rangle\epsilon_{ch}
\langle N\rangle,
\end{equation}
\begin{equation}
f_{0,1}(ch, \gamma) =\langle n_{\gamma}\rangle=\langle
f\rangle(\epsilon_{1} +2\epsilon_{2})\langle N\rangle,
\end{equation}
\begin{equation}
f_{2,0}(ch, \gamma) =\langle
n_{ch}(n_{ch}-1)\rangle=\langle(1-f)^{2}\rangle\epsilon_{ch}^{2}\langle
N(N-1)\rangle,
\end{equation}
\begin{equation}
f_{1,1}(ch, \gamma) =\langle n_{ch}n_{\gamma}\rangle = \langle
f(1-f)\rangle\epsilon_{ch} (\epsilon_{1} +2\epsilon_{2})\langle
N(N-1)\rangle,
\end{equation}
\begin{equation}
f_{0,2}(ch, \gamma) =\langle n_{\gamma}(n_{\gamma}-1) \rangle=\langle
f^{2}\rangle(\epsilon_{1} +2\epsilon_{2})^{2}\langle N(N-1)\rangle +
2\epsilon_{2} \langle f\rangle\langle N\rangle.
\end{equation}
In Eqs. (25)-(29) the overall statistical averages for the charged, photon,
and charged-photon factorial
moments are expressed, in an obvious notation, in terms of the independent
moments taken
with respect to
 the $P(N)$ and $p(f)$ distributions.

Finally, we turn to the effect of the MiniMax trigger on these
considerations. The MiniMax trigger
requires, among other things, a coincidence in the signals from
scintillator counters located behind 
both the converter and the entire tracking telescope. In consequence,
events in which no charged 
particle or converted $\gamma$ goes through the acceptance of the detector
are triggered with 
different (and lower) efficiency, $\epsilon$, than events in which either a
charged particle or 
$\gamma$ conversion products go through the aperture. An effective model
for the effect of the 
MiniMax trigger on the probability, $p^{obs}(n_{ch},n_{\gamma})$, for
observing an event 
with $n_{ch}$ charged
particles and $n_{\gamma}$ converted $\gamma$'s passing through the
acceptance is given by the 
proportionalities
\begin{equation}
p^{trig}(0,0) = \epsilon \alpha p^{obs}(0,0), \; \; \; \; n_{ch}=n_{\gamma}=0,
\end{equation}
and
\begin{equation}
p^{trig}(n_{ch},n_{\gamma}) = \alpha p^{obs}(n_{ch},n_{\gamma}), \; \; \;
\; n_{ch}+ n_{\gamma}>0.
\end{equation}
Here $p^{trig}(n_{ch},n_{\gamma})$ is the measured probability of seeing an
event, including 
the effects of both the trigger and the 
particle detection efficiencies, while $p^{obs}(n_{ch},n_{\gamma})$
presumes perfect triggering. 
If 
\begin{equation}
\alpha = [1+ (1-\epsilon)p^{obs}(0,0)]^{-1},
\end{equation}
$p^{trig}$ will be properly normalized if $p^{obs}$ is.

The bivariate factorial moments 
transform homogeneously under the transformation (30)-(32) incorporating
differential 
trigger efficiencies,
\begin{equation}
f_{i,j}(ch, \gamma) \rightarrow \alpha f_{i,j}(ch, \gamma).
\end{equation}

\section{ROBUST OBSERVABLES}
The second-order factorial moments (25)-(29) represent the lowest-order
correlative effects
among charged pions and photons. We see from (29) that the gamma-gamma correlations are 
distinguished by the term $2\epsilon_{2} \langle f\rangle\langle N\rangle$
for observing the two photons from a single neutral 
pion, so that 
this average will not be a component of a robust measure involving only
first and second order
moments. This suggests the construction of
a measure from the moments (25)-(28) in the form of a ratio in order to
cancel out as many effects
as possible, apart from the $p(f)$ averages, that reflect the particular
details of the 
production mechanism. 

Consider, then, the ratio
\begin{equation}
  r_{1,1}= {{\langle n_{ch}n_{\gamma}\rangle\langle n_{ch}\rangle } \over
{\langle n_{ch}(n_{ch}-1)\rangle\langle n_{\gamma}\rangle}}.
 \end{equation}
For generating functions of the form (23), we find from (25)-(28) that
\begin{equation}
  r_{1,1}= {{\langle f(1-f)\rangle\langle (1-f)\rangle } \over {\langle
(1-f)^{2}\rangle\langle
 f\rangle}}, 
 \end{equation}
an expression in which all reference to the background distribution $P(N)$
and the efficiencies
$\epsilon_{1}$, $\epsilon_{2}$, and $\epsilon_{ch}$ have cancelled out.
 Further, we see that
\begin{equation}
r_{1,1}  \rightarrow r_{1,1}
\end{equation}
under the transformation (30)-(32) so that $r_{1,1}$ is  a ``robust
observable" in the sense referred
to in Sec. 1.

It follows from (35) that
\begin{equation}
 r_{1,1} \leq 1,
\end{equation}
where the equality is realized for generic pion production,
$p(f)=\delta(f-\hat{f})$,
\begin{equation}
  r_{1,1}(generic)= 1, 
 \end{equation}
independently of $\hat{f}$. The realization of the limit (38) in the UA5
data at 200 GeV and 900 GeV [25], 
and in Monte Carlo simulations at 1.8 TeV, both of which include nonpionic 
sources of charged particles and photons, is considered in Sec. 6.

For a DCC distribution, $p(f)=1/(2\sqrt {f})$, one finds
\begin{equation}
  r_{1,1}(DCC)= {1 \over 2}.
 \end{equation}
This clearly distinguishes the pure DCC and generic distributions.

The values (38) and (39) represent the limiting extremes of a mixture of
generic and DCC 
distributions. Generally, broad (DCC) and narrow (generic)
statistical distributions can be distinguished in a mixture of the two by
means of higher-order 
moments that are 
sensitive to the tail of the charged-particle/photon distribution. Robust
combinations of
 these higher-order moments that are generalizations of $r_{1,1}$ 
will be of greatest practical value in an analysis of data in which a 
discernable fraction of DCC form is expected to appear.

Let us first note that the normalized factorial moments 
\begin{equation}
F_{i} \equiv \frac{\langle N(N-1)\ldots (N-i+1)\rangle} {\langle N\rangle ^{i}} 
\end{equation}
are unity if the parent distribution $P(N)$ is Poisson. Therefore,
deviations from purely
 random fluctuations
are measured by the departure of the $F_{i}$'s from unity. A bivariate
generalization
 of the $F_{i}$'s is given by 
\begin{equation}
 F_{i,j}=\frac{\left< n_{ch}(n_{ch}-1)\ldots
(n_{ch}-i+1)~n_{\gamma}(n_{\gamma}-1) \ldots (n_{\gamma}-j+1)\right>}
{\left<n_{ch}\right>^{i}\left<n_{\gamma}\right>^{j}}\ .
\end{equation}
In particular, one finds that 
\begin{equation}
F_{i,0} = \frac{F_{i}\left<(1-f)^{i}\right>} {\left<(1-f)\right>^{i}}
\end{equation}
and
\begin{equation}
F_{i,1} = \frac{F_{i+1}\left<f(1-f)^{i}\right>}
{\left<f\right>\left<(1-f)\right>^{i}},
\end{equation} 
where $F_{i}$ refers to the $i$th normalized factorial moment (40) of the
$P(N)$ distribution 
for the total multiplicity. We note that
\begin{equation}
F_{i,j} \rightarrow  \alpha^{1-i-j}F_{i,j}
\end{equation}
under the transformation (30)-(32).

Evidently, $r_{1,1}= F_{1,1}/F_{2,0}$. From (42) and (43) we find a
generalization
 of $r_{1,1}$ to a family, $R$, of 
robust observables:
\begin{equation}
r_{i,1}=\frac{F_{i,1}}{F_{i+1,0}} 
= \frac{\left<(1-f)\right>\left<f(1-f)^{i}\right>}{\left<f\right>
\left<(1-f)^{i+1}\right>} .
\end{equation}
Moreover, one finds that for all $i \geq 1$
\begin{eqnarray} 
r_{i,1}(generic) &=& 1, \nonumber \\
r_{i,1}(DCC) &=& \frac{1}{i+1}  ,
\end{eqnarray}
in the two cases.
Thus, $r_{i,1}$ becomes more sensitive to the difference between DCC and
generic production 
mechanisms with increasing order of the moments. This reflects the
broadness characteristic of the 
DCC distribution in the neutral fraction $f$ compared to the generic case.

The ratios
\begin{equation}
r_{i,j}=\frac{F_{i,j}}{F_{i+j,0}} 
\end{equation} 
are not robust because the moments $F_{i,j}$ for arbitrary $i$ and $j$ are
not independent
of the photon detection efficiencies. However,
the terms involving these efficiencies can be expessed in terms of only one
combination of these 
parameters, namely
\begin{equation}
\xi =
\frac{2\epsilon_{2}}{(\epsilon_{1}+2\epsilon_{2})\left<n_{\gamma}\right>}, 
\end{equation}
along with the mean number of photons, as 
\begin{equation}
F_{i,j} = \sum_{m=0}^{[j/2]}c_{j,m}\xi^{m}F_{i+j-m}
\frac{\left<(1-f)^{i}f^{j-m}\right>}{\left<(1-f)\right>^{i}
\left<f\right>^{j-m}}, 
\end{equation} 
The coefficients $c_{j,m}$ are obtained from the identity, true for any
differentiable function, $D(z^{2})$,
\begin{equation}
\frac{d^{j}D(z^{2})}{(dz)^{j}} = \sum_{m=0}^{[j/2]}c_{j,m}2^{m}(2z)^{j-2m}
\frac{d^{j-m}D(z^{2})}{(dz^{2})^{j-m}}
\end{equation}
The first few $c_{j,m}$ are [28]:
\begin{eqnarray}
c_{j,0}& = & 1, \nonumber \\
c_{j,1}& = & j(j-1)/2, \nonumber \\
c_{j,2}& = & 3~j!/4!(j-4)!. \
\end{eqnarray}

One can use the ratios $r_{i,j}$'s in the analysis
of experimental distributions, with the understanding that the parameter
$\xi$ is to be 
determined from the data. Generally, we have the bounds and limiting values
\begin{equation}
r_{i,j}(generic) \geq 1,
\end{equation}
\begin{equation}
\left( r_{i,j}(generic)  \right)_{\xi=0}= 1,
\end{equation}
and
\begin{equation}
\left( r_{i,j}(DCC)  \right)_{\xi=0} ={{i!(2j-1)!!}\over {i+j}}.
\end{equation}

\section{SENSITIVITY TO DCC ADMIXTURES}
We next turn to the question of what can be said about robust observables
when there is 
an 
admixture of DCC and generic multipion production. There is considerable 
theoretical uncertainty about how such an admixture would arise in hadronic
collisions
and so there are many possibilities for extending the development given in
the preceding 
sections. Our objective in this section is only to 
provide a formalism in which the sensitivity of experimental results to the 
presence of DCC or some other 
anomalous mechanism can be investigated. Thus, it will suffice to address
this question only in 
the context of a 
few simple limiting models of 
 pion production containing both generic and DCC components. Specifically,
we consider modifications of the generating-function formalism we have
developed 
in the preceding sections
 in 
three different 
scenarios for mixing DCC and generic multiparticle production.  
Then we examine the impact of these modifications on the values of the
robust observables.

\subsection{Exclusive Production}
First, let us consider the possibility of what we refer to as {\em
exclusive} production. 
That is, in a given event, particle production is either the result of the
formation of a 
DCC with probability $\lambda$, or it is generic with binomially
distributed charged and 
neutral particles with probability $1- \lambda$. The picture of exclusive
production could be 
regarded as a first-order phenomenology of very high-energy cosmic-ray 
interactions, which
seem to  divide themselves into what appear to be generic and anomalous
classes [29].

The generating function for the 
exclusive 
production of charged pions and the photons resulting from $\pi^{0}$ decay
is simply the 
weighted sum of the generic and DCC generating functions:   
 \begin{equation}
G_{excl}(z_{ch}, z_{\gamma}, \lambda) = (1- \lambda)G_{ generic}(z_{ch},
z_{\gamma}) + \lambda G_{DCC}(z_{ch}, z_{\gamma}). 
\end{equation}
Here $G_{ generic}(z_{ch}, z_{\gamma})$ and $G_{DCC}(z_{ch}, z_{\gamma})$
are obtained from
 (23) for the cases $p(f)= \delta(f - \hat{f})$ and $p(f)= 1/(2\sqrt{f})$, 
respectively, and where the distributions $P(N)$ of the total number of
pions are generally 
different 
in the two cases. 

The expressions for the moments $r_{i,1}$ obtained using 
$G_{excl}(z_{ch}, z_{\gamma}, \lambda)$ interpolate between the generic and
DCC limits as 
$\lambda$ varies between 0 and 1. For example, since
\begin{eqnarray}
f^{excl}_{i,j} &=&(1-\lambda)f^{gen}_{i,j} +\lambda f^{DCC}_{i,j} \nonumber \\
               &=&f^{gen}_{i,j}( 1+\lambda({f^{DCC}_{i,j} \over
f^{gen}_{i,j}}-1)), 
\end{eqnarray} 
it follows, using the results of Section 3, that one can write
\begin{equation}
r^{excl}_{1,1}(\lambda)=
{ [1+ \lambda({2\over 15 {\hat f}(1-{\hat f})} {\langle N(N-1)\rangle^{DCC}
\over \langle N(N-1)\rangle^{Gen}} -1)] [1+ \lambda({2\over 3 (1-{\hat f})}
{\langle N)\rangle^{DCC} \over \langle N\rangle^{Gen}} -1)] \over [1+
\lambda({8\over 15 (1-{\hat f})^2}
{\langle N(N-1)\rangle^{DCC} \over \langle N(N-1)\rangle^{Gen}} -1)][1+
\lambda({1\over 3 {\hat f}}
{\langle N)\rangle^{DCC} \over \langle N\rangle^{Gen}} -1)] } 
\end{equation}
Note that this expression explicitly depends on the relative size of the
DCC and the 
generic factorial moments. Technically, this ratio
 is no longer ``robust" in the sense of the preceding section. However, it
still does not depend 
upon efficiency corrections. In addition, the extra dependence will be an
{\em advantage}
 if DCC dominates the high-multiplicity tail of the distribution.

\subsection{Independent Production}
A second possible production scenario is where the occurrence of DCC in an
event is 
independent 
of the pions that are produced generically. Independent production implies
that the 
probability
$P_{DCC}(N)$ for producing $N$ DCC pions is independent of the probability 
$P_{generic}(N)$ for producing $N$ binomially distributed pions, so that 
the generating function 
factors into a product,
\begin{equation}
G_{ind}(z_{ch}, z_{\gamma}) = G_{ generic}(z_{ch},
z_{\gamma})G_{DCC}(z_{ch}, z_{\gamma}). 
\end{equation}
Thus, we find
\begin{equation}
f^{ind}_{i,j} = \sum_{\alpha=0}^i \sum_{\beta=0}^j \left( \begin{array}{c} 
i\\
\alpha
\end{array}\right)\left( \begin{array}{c} 
j\\
\beta
\end{array}\right) f^{Gen}_{i-\alpha,j-\beta} f^{DCC}_{\alpha,\beta}. 
\end{equation}
Hence, using the results of the previous sections, it follows that, for example,
\begin{equation}
r^{ind}_{1,1} =
{ [1+ {\langle N \rangle^{Gen} \langle N \rangle^{DCC} \over 
\langle N(N-1) \rangle^{Gen}}
( {2\over 3(1-{\hat f})} + {1\over 3{\hat f}}) + { 2 \langle
N(N-1)\rangle^{DCC}\over
15 {\hat f} (1-{\hat f}) \langle N(N-1)\rangle^{Gen}} ] [1 + {2 \langle N
\rangle^{DCC} \over
3 (1-{\hat f}) \langle N \rangle^{Gen}}] \over [1+ {\langle N \rangle^{Gen}
\langle N \rangle^{DCC} \over 
\langle N(N-1) \rangle^{Gen}}
( {4\over 3(1-{\hat f})})
+ {8 \langle N(N-1)\rangle^{DCC}\over
15 (1-{\hat f})^2 \langle N(N-1)\rangle^{Gen}} ] [1 + {1 \langle N
\rangle^{DCC} \over
3 {\hat f} \langle N \rangle^{Gen}}] }.
\end{equation}
Again the sensitivity to the independent production of DCC is dependent on
the ratios of 
DCC and generic factorial moments, but not to the efficiency corrections.

We note that in the independent production model
\begin{equation}
\ln G_{ind}(z_{ch}, z_{\gamma}) = \ln G_{ generic}(z_{ch}, z_{\gamma}) +\ln
G_{DCC}(z_{ch}, z_{\gamma}), 
\end{equation}  
which suggests an analysis in terms of a bivariate generalization of
single-variable
 cumulant moments [16, 18-20].  
We define bivariate cumulants for $i+j > 0$ by
\begin{equation}
k_{i,j} = \left( \frac{\partial^{i+j}}{\partial z_{ch}^{i}\partial
z_{\gamma}^{j}}\ln G\right)_{z_{ch}=z_{\gamma}=1}. 
\end{equation}
>From (58) we see that in this production scenario, the cumulants are additive:
\begin{equation}
k_{i,j}^{ind} = k_{i,j}^{generic} + k_{i,j}^{DCC}. 
\end{equation}
For single-variable probability distributions, cumulants reflect non-random
correlations 
in that they vanish for a Poisson distribution. In the bivariate case their
properties 
as a measure of correlations are not so direct.

As with the bivariate normalized factorial moments (41), we introduce
normalized 
bivariate cumulant moments:
\begin{equation}
K_{i,j} = \left<n_{ch}\right>^{-i}\left<n_{\gamma}\right>^{-j}\left(
\frac{\partial^{i+j}}{\partial z_{ch}^{i}\partial z_{\gamma}^{j}}\ln
G\right)_{z_{ch},z_{\gamma}=1} . 
\end{equation}
In the independent model we obtain for $K_{i,j}^{ind}$ the weighted sum
\begin{equation}
K_{i,j}^{ind} = \lambda_{ch}^{i}\lambda_{\gamma}^{j}K_{i,j}^{DCC} +
(1-\lambda_{ch})^{i}(1-\lambda_{\gamma})^{j}K_{i,j}^{generic},
\end{equation}
where
\begin{equation}
\lambda_{ch,\gamma} = \frac{\left<n_{ch,\gamma}\right>_{DCC}}
{\left<n_{ch,\gamma}\right>}
\end{equation}
are the fractions of the mean charged or photon multiplicities attributed
to the DCC.

The formulae for the normalized cumulant moments for DCC and generic
subsamples are 
obtained in a straightforward manner. As before, most of the efficiency
corrections cancel out. 
However, the cumulant moments do not scale 
homogeneously under the differential trigger inefficiency characteristic of
MiniMax. While 
this is disadvantageous for the early MiniMax analyses, there is reason to 
expect that they will be eventually of substantial utility in MiniMax as
well as in other experiments.

\subsection{Associated Production}
A third possiblity for the contamination of a DCC signal by generic
multiparticle production 
is what can be called associated production. For example, in the Baked
Alaska model [8] the 
number of DCC pions is estimated to scale as
\begin{equation}
N_{DCC} \sim (N_{generic})^{3/2}.
\end{equation}
A simpler case, which is also a credible scenario, is where the amount of
DCC production 
is, on the average, proportional to the 
amount of generic production. It then follows using the cluster theorem
[27], that
\begin{eqnarray}
G_{assoc}(z_{ch}, z_{\gamma}; \lambda) & = & \int_{0}^{1}df_{b} \,
p_{b}(f_{b})\int_{0}^{1}df_{d} \, p_{d}(f_{d}) \nonumber \\
                                       &   &
\sum_{N=0}^{\infty}P(N)[(1-\lambda)g_{b}(z_{ch}, z_{\gamma}) + \lambda
g_{d}(z_{ch}, z_{\gamma})]^{N},
\end{eqnarray}
where, 
\begin{equation}
g_{A}(z_{ch}, z_{\gamma}) = f_{A}g_{0}(z_{\gamma}) + (1-f_{A})g_{ch}(z_{ch}),
\end{equation}
\begin{equation}
p_{b}(f_{b}) =\delta (f_{b} - \hat{f}),
\end{equation} 
\begin{equation}
p_{d}(f_{d}) = 1/(2\sqrt{f_{d}}),
\end{equation}
and the index $A$ takes the values $b$ and $d$ in the binomial and DCC
cases, respectively.
 As before, 
one can carry out the calculation of the robust observables which results
in formulae that 
interpolate as a function of the fraction, $\lambda$, of DCC admixture
between the generic and 
DCC limits. Note that in this case there would be only a single parent
$P(N)$, common to both the 
generic and 
DCC production. Using the results of the previous section, one can
calculate, for example, 
\begin{equation}
r^{assoc}_{1,1}(\lambda)=
{[(1-\lambda)^{2}{\hat f}(1-{\hat f}) +
{1\over 3}\lambda (1-\lambda) (1 +{\hat f}) +{2\over 15} \lambda^2]
[(1-\lambda)(1-{\hat f}) +{2\over 3}\lambda ]\over [(1-\lambda)^{2}(1-{\hat
f})^2 +
{4\over 3}\lambda (1-\lambda) (1 -{\hat f}) +{8\over 15}
\lambda^2][(1-\lambda){\hat f} +{1\over 3}\lambda ]},
\end{equation}
which, in contrast to the other two cases, (57) and (60), is a fully robust
observable.

\subsection{Other Particles}
A similar framework can be used to discuss the sensitivity of the
predictions to the production of 
particles other than pions. This is of potential concern, since $K$ and
$\eta^{0}$ production may be 
a substantial fraction of pion production [17, 25, 30]. In particular, the
$\eta^{0}/\pi^{0}$ ratio 
can be quite large leading to an excess of gammas over the case of pions alone, 
where $\langle n_{\gamma} \rangle = \langle n_{ch} \rangle$.

Relatively little is known about $K$ and $\eta^{0}$ distributions at the
highest energies, especially in 
forward directions, so, while an 
independent production model might be more accurate, we will limit our
considerations at the moment 
to the context of an ``associated" production model. In essence, we are
thus assuming that a system 
of parent partons is created in the collision process, and that this system
then evolves into a system
of $N$ hadrons with probability $P(N)$, with the hadrons independently
partitioned into various species. 

Let the index $i$ run over the various types of hadrons that are produced.
The $i'th$ type of hadron 
is produced with relative probability $\lambda_{i}$ (with
$\sum_{i}\lambda_{i} =1$). These hadrons then decay into charged 
particles and $\gamma$'s, and each species of hadron is characterized by a
generating function for 
detecting the products of that species,
\begin{equation}
g_{i}(z_{ch}, z_{\gamma}) = \sum_{n_{ch}}\sum_{n_{\gamma}}
\epsilon^{(i)}_{n_{ch},n_{\gamma}}z_{ch}^{n_{ch}} z_{\gamma}^{n_{\gamma}}, 
\end{equation}
where $g_{i}(1,1)=1$. Then the observed generating function, neglecting DCC
production, can be written as
\begin{equation}
G_{obs}(z_{ch}, z_{\gamma}) = \sum_{N} P(N)[\sum_{i}\lambda_{i}
g_{i}(z_{ch}, z_{\gamma})]^{N}.     
\end{equation}

We can now make a few observations about the impact of contamination of the
predictions that 
arise from 
$K$ and $\eta^{0}$ production. The following estimates of the effects of
various particle types on
the magnitude of 
$r_{1,1}$ draw upon the simulations specific to MiniMax reported in Section 6.

First, we note that the $K^{\pm}$'s, which are seen simply as charged
particles in MiniMax, appear just as 
another source of charged particles from the collision point and so modify
the neutral fraction, but 
are otherwise benign. Similarly, the $K_{L}$'s have a decay length much
longer on average than the length 
of the MiniMax detector. In consequence they are only detected, but not
identified, when they 
interact strongly in the converter used to identify photons. On the
relatively rare occasions when 
$K_{L}$'s do interact in the converter, they are misidentified as
$\gamma$'s. This will also 
influence the net 
neutral fraction that is observed, but is also otherwise benign. In
conclusion, the 
associated production of $K^{\pm}$'s and $K_{L}$'s will not change the
values of the $r_{i,j}$'s 
predicted in Section 4 for ``generic" production.

The case of $K_{S}$ production is rather interesting since the $K_{S}$
decay modes,
 $K_{S} \to \pi^+ \pi^- (69\%)$, $K_{S} \to \pi^0 \pi^0 (31\%)$, are
essentially those of an 
isosinglet DCC with one pair of pions. That is, in regard to the statistics
of the 
particles produced, $K_{S}$ decays are essentially identical to those of
the smallest conceivable 
domain of DCC's. As such, $K_{S}$ production is in principle of interest
from the point of sensitivity to very small domains of DCC. 

Let us consider associated 
production of $K_{S}$'s with fraction $\lambda_{K_{S}}$. The generating
function for studying 
the modification of generic production is thus
\begin{equation}
G_{K_{S}}(z_c,z_\gamma;\lambda_{K_{S}})=
\sum_N P(N) [ (1-\lambda_{K_{S}})g_{gen}(z_c,z_\gamma) + \lambda_{K_{S}}
g_{K_{S}}(z_c,z_\gamma)]^N \,.
\end{equation}
Using previous methods, one finds that 
\begin{equation}
(r_{1,1}^{K_{S}}(\lambda))^{-1}=
1+ {2 \langle N \rangle \lambda_{K_{S}} \epsilon^{K_{S}}_{2,0}\over \langle
N(N-1)\rangle ((1-\lambda_{K_{S}})(1-{\hat f}) \epsilon_{ch} +
\lambda_{K_{S}}(\epsilon^{K_{S}}_{1,0}+2\epsilon^{K_{S}}_{2,0}))^2}
\end{equation}
which is manifestly not robust.

$K_{S}$'s are not DCC domains; they are, rather, particles of well-defined
mass and 
a lifetime such that most of them have decayed before reaching the 
MiniMax detector, and their decay products have strong correlations and are
not vertexed to the 
collison point. As a consequence, in MiniMax the acceptance for 2 charged
pions from 
a single $K_{S}$ is about 4\%. Consequently, the impact of $K_{S}$
production on the 
MiniMax systematics is expected to be quite small. 

One can similarly study the impact of $\eta^{0}$ production on the
idealized predictions of Section 4. 
The $\eta^{0}$ has a wider variety of decay modes and all of the charged
particles and 
$\gamma$'s from the decays are collision vertexed. Thus
$g_{\eta^{0}}(z_c,z_\gamma)$ is more 
complicated, 
but the calculations follow closely those outlined for $K_{S}$ decays. In addition
to having decay modes with more than a single charged particle, there are
decay modes 
with intrinsic charged-$\gamma$ correlations, as well as the 
charged-charged correlations which entered into the $K_{S}$ analysis. The
conclusion is, nevertheless,
much the same.

\subsection{Detector Effects}
Finally, we note that the formalism we have developed can be extended to 
consider contamination due to detector-related effects. For example, in
detectors which identify 
gamma rays by electromagnetic calorimetry, charged hadrons can also be
identified as photons 
when they interact strongly in the calorimeter. 
For example, in WA98 [31], a heavy-ion 
experiment at CERN which has instituted a DCC search, this is expected to
occur approximately 20\% of the time.
Such misidentifications can be handled by using an 
appropriate form of the generating function $g_{i}(z_{ch},z_{\gamma})$. 
For example, 
\begin{equation}
g_{\pi^\pm}(z_{ch},z_{\gamma}) =
\epsilon^{\pi^\pm}_{0,0}+\epsilon^{\pi^\pm}_{1,0}z_{ch} +
\epsilon^{\pi^\pm}_{1,1}z_{ch} z_\gamma
\end{equation}
would be suitable if some fraction $\epsilon^{\pi^\pm}_{11}$ of the charged
pions were 
tagged as both charged particles and photons because of the calorimeter's
response.

\section{ROBUST OBSERVABLES IN PRACTICE} 

We now turn to the utilization of the robust observables for analyzing collider 
data, both actual or simulated. 
As we saw in the last section, the assumptions made earlier
 are idealizations that are violated by some types of production mechanisms 
and by less than ideal detector performance. In this section we examine the
properties of the   
robust moments in the context of the UA5 data and 
Monte-Carlo simulations of the MiniMax detector in order to assess the
importance of these 
violations in practice.

\subsection{$r_{1,1}$ from UA5}
For collider energies of 200 GeV and 900 GeV, UA5 measured the inclusive
charged-particle 
and photon
$dN/d\eta$ distributions, as well as the corresponding charged-charged and the 
charged-photon correlation functions, $C_{ch,ch}(\eta_{ch}=0, \eta_{ch})$ 
and $C_{\gamma,ch}(\eta_{\gamma}=0, \eta_{ch})$, respectively [25]. Here,
$\eta_{ch}$ and 
$\eta_{\gamma}$
denote the charged-particle and photon pseudorapidities, respectively. The
 measurements were carried out over about four units of $|\eta_{ch}|$. The mean 
values $\langle n_{ch} \rangle$ and $\langle n_{\gamma} \rangle$
can be calculated for different pseudorapidity bins 
using the experimental $dN/d\eta$ distributions. Under 
the assumption that $C_{ch,ch}(\eta_{1}, \eta_{2})$ and 
$C_{\gamma,ch}(\eta_{1}, \eta_{2})$ depend only on the absolute value 
of $|\eta_{1}-\eta_{2}|$, the second-order moments that enter 
into $r_{1,1}$ can be also be calculated for corresponding pseudorapidity 
bins. Despite large uncertainties in the UA5 photon data and the validity
of our 
assumptions about the correlation functions, we 
find $r_{1,1} = 1.0 \pm 0.10$ for the different energies and various bin
choices. 

\subsection{Simulations}

While we believe the robust observables will find general 
application in experimental searches for DCC, we are motivated here primarily 
by the MiniMax experimental situation. In this context, in order to make a
rough 
check of the validity of the assumptions we have made in the opening sections,
 we next describe a series of complete simulations of the MiniMax experiment. 

Minimum bias 
events are generated in PYTHIA version 5.702 and JETSET 7.401 [32]. 
The output of PYTHIA is then used as input to the 
simulation of the detector response using GEANT, version 3.21 [33]. The GEANT 
output is then 
put through a full tracking and analysis chain. The resulting frequency 
distributions for observing $n_{ch}$ charged tracks and $n_{\gamma}$ converted 
photons are then used to calculate the various robust observables. 
Similar studies, in which the output of PYTHIA is replaced or augmented 
by the output of a DCC generator, are also carried out. We find the results 
of these simulations to be in agreement with expectations from our calculations
 in the previous sections. 

\subsubsection{Standard Monte Carlo}
PYTHIA is used to simulate the minimum bias collisions at $\sqrt{s}=1.8$
TeV. Default values 
are taken for all 
parameters except that particles with a mean decay length greater 
than 1 cm were not allowed to decay.

There are no published data on multiparticle production at 1.8 TeV in the 
pseudorapidity interval covered by MiniMax, so there is no independent
check on the accuracy of the
simulations. For recent measurements at 630 GeV [34], the 
agreement between PYTHIA and the
 $dN/d\eta$ data, 
in a range of pseudorapidity including that of MiniMax, is less than ideal. 
Nonetheless, the PYTHIA output represents a useful benchmark.

The particles generated in a simulated collision are then taken as input
into a GEANT
simulation of the 
detector and its environment. The experimental data give evidence of a
large background of 
particles aising from interactions in
 material immediately surrounding the detector. Therefore,  many nearby
objects are included in the 
simulation. GEANT propagates the particles through the detector and its
surroundings and produces
a simulation of the data that are produced by the actual detector. Despite
care in including all 
relevant aspects of the detector and its environment, the GEANT data show a
smaller number of 
reporting wires in the MWPC's than do the actual data by a factor of two.

GEANT data are written to a file that is used as input to the same code
that is used for the analysis 
of the actual MiniMax data. The analysis proceeds in two stages.
First, a tracking code is used to find track segments in front of (heads)
and behind (tails)
 the converter plane. The output of this calculation is used by a second
code (vertexer) that 
determines the number of charged particles and $\gamma$'s observed in the
event. In so doing, it 
counts a charged track to be a head that can be joined to at least one tail. 
A $\gamma$ conversion is taken to be one or more tails emanating from the
same point in the converter 
without an accompanying head. Candidate charged and $\gamma$ tracks are
required to point to 
within some given distance from the collision point in order to remove
secondary particles from 
material adjacent to the detector and fake tracks arising from chance
combinations of random 
reporting wires. The parameters used in the vertexer are determined by
optimizing the reconstruction 
of the events generated by PYTHIA and GEANT.

This track-reconstruction procedure is still under development.
 It does not satisfy all of the 
assumptions made in Section 1 regarding tracking efficiency. In particular,
the reconstruction 
efficiency may depend on the multiplicity and proximity of tracks. 

\subsubsection{DCC generator}
DCC production is modelled according to the $1/(2\sqrt{f})$ distribution. 
For the present simulation, the DCC domain size in $\eta - \phi$ space 
is taken to be
 on the order 
of the detector acceptance. The c.m. momentum of the DCC is directed 
at the center of the acceptance with a reasonably large $p_{T}$. 
We assume that the number of pions in the DCC is independent of the 
central pseudorapidity of the DCC. The 
ratio of the mean energy density of DCC pions to that of generically 
produced pions is then approximately constant; we 
take the ratio to be unity. 

DCC's are generated using what could be called a
 ``snowball" model in reference to the low pion momenta in the DCC c.m. 
The number, $N_{DCC}$, of DCC pions is 
chosen using a Poission distribution with mean $\mu_{DCC}$.

The neutral fraction is
generated using the 
transformation method, where, if $x$ is a uniform deviate, $f=x^2$ is
distributed 
according to $1/(2\sqrt{f})$. A uniform deviate $x_{i}$ is then generated
for each of the $N_{DCC}$ pions; if $x_{i} < f$, the pion is defined to be 
neutral, otherwise it is defined to be charged. This procedure implements the 
$1/(2\sqrt{f})$ distribution exactly; if one takes the viewpoint that the 
isosinglet distribution is more fundamental, then this procedure can be 
viewed as an approximation to it which is valid in the limit that the 
total number of pions is large, and one is sampling a subset of the DCC. 
The actual distribution is, of course, an experimental question.

Each of the pions is assigned a 3-momentum in the DCC c.m. system by drawing 
from a zero-mean Gaussian distribution with a 
variance $\langle {\vec p}\cdot {\vec p}\rangle =3 \sigma_p^2$. 

The DCC is then boosted such that the momentum of the DCC c.m. is in the
direction of 
the center of the MiniMax detector at $\eta =4.1$, and so that the DCC pions 
have $\langle p_{T} \rangle \sim \sigma_{p}$. If the pions are not too
relativistic 
in the DCC c.m.
 frame, the boosted DCC domain is approximately 
circular in $\eta-\phi$ space, with radius $R_{DCC}\sim \sigma_{p}/p_{T}$. 

The results we report next are based on Monte-Carlo simulations in 
which $\sigma_p=0.1$ GeV and $p_{T} =0.14$ GeV; hence $R_{DCC}\sim 0.7$, 
the typical radius of a hadronic jet.
The Poisson mean for the number of DCC pions has the value $\mu_{DCC}=5.5$,
which
 corresponds to an energy density in lego space comparable to 
that of generic production.
The Monte-Carlo simulation of DCC production is used to generate pure DCC 
events. These events are then run through the same GEANT simulation as the 
PYTHIA events, except that the trigger is not used since no 
particles go in the $\bar{p}$ direction.

\subsection{Results}

Once the number of charged tracks and $\gamma$'s passing into the acceptance is 
determined, the moments and $r_{ij}$ are calculated. Statistical 
errors are estimated assuming Poisson fluctuations and the standard 
propagation of errors formalism [35].

The results obtained for approximately $5\times 10^4$ PYTHIA events and 
$2\times 10^4$ pure DCC events are shown in Table 1. For purposes of
comparison, 
the predicted values for idealized binomial and DCC distributions are 
included. For those ratios involving higher-order moments of the number of 
observed, converted $\gamma$'s, the predictions are nonrobust, as discussed in 
Section 4, and
depend on $\xi$, which is determined from the relationship between 
$f_{0,2}$, $f_{2,0}$, and $<\!n_\gamma\!>$, assuming a binomial distribution. 
The same values are used in correcting the DCC predictions for the higher 
order moments. In particular,
it is assumed that  
$2 \epsilon_2/(\epsilon_1 + 2\epsilon_2) \approx 0.08 \pm 0.01$ obtained
from PYTHIA for 
generic production
has the same value for DCC production. This is 
certainly violated in practice, for 
the simulated DCC pions have significantly lower $\langle p_{T} \rangle
\sim \sigma_{p}$
 than those generated by PYTHIA, and hence the 
probablility of both $\gamma$'s from a $\pi^{0}$ decay being in the acceptance, 
which is reflected in $\epsilon_2$, will be different. In addition, the
$F_{i}$'s are 
also taken to be the same in the DCC case as in the PYTHIA case, which is also 
clearly a poor assumption. We have chosen to display the data in the 
manner shown, however, in order to illustrate the problems 
which will arise in DCC searches using these moments.

There is general agreement between the ``predictions" 
based on the analysis in Section 4 of this paper, and the results of 
these full simulations. One of the striking features of these results is
how well the 
PYTHIA/GEANT simulation, which includes resonance production, simulations of 
detector effects, among other features, matches the predictions of a simple
binomial model.

In order to illustrate the effect of an admixture of DCC with generic 
events, where the amount of DCC produced is independent of the amount of 
generic production, DCC domains from the DCC-generator/GEANT are added to 
various fractions of random PYTHIA/GEANT events. This represents a mixture
of the independent 
and exclusive models considered in Section 5. The effect on the 
$r_{i,1}$ is shown in Table 2.

These simulations support 
the expectation that the robust observables introduced in this paper will
be a useful 
analysis tool, even though all of the technical 
requirements of robustness may not be met. Thus, these observables provide a well-defined
 framework for describing correlations in such way
 that many systematic uncertainties cancel out.

\section{CONCLUDING REMARKS}
Most of the experimental analyses and 
theoretical studies of multihadron production 
have concentrated only on charged-hadron production, for which the bulk of
the data have been taken; for exceptions to this, see [21-25]. The 
questions we have addressed concerning the neutral-hadron component of
multiparticle production 
have received little attention, but are vital for our MiniMax experiment.

The robust observables $R$ which are here proposed appear, on the basis of
the analytic 
calculations 
and Monte Carlo simulations we have presented, to be of considerable value
in all future 
analyses of
 charged-particle/photon distributions in high-energy hadron and heavy-ion
collisions, and 
especially with respect to the search for disoriented chiral condensate.

While these observables are manifestly robust, there are still clear
limitations to their use 
which must be eventually be addressed. We have said little about
momentum-dependent 
efficiencies; this
will, at the formal level, require generating functions to be generalized
to generating 
functionals [22-24]. 
At this level, even the choice of the parent generating functional may have
considerable ambiguity 
due to a lack of consensus on the underlying physics; e.g., can the
Poisson-transform structure 
of Eq. (6) be simply generalized?

At a more practical level, the issue of correlated efficiencies, especially
with respect to total 
multiplicity and background level, is vital. Here the features of the
individual experiment 
and its 
environment are essential, and a strong interplay between simulations and
the analysis of 
real data 
is required.

Finally, in experiments with large acceptance, even for pure DCC production
the chiral order 
parameter may be different in different portions of the $\eta-\phi$, or
lego, phase space. In this case 
the formalism 
we have presented must undergo further generalization.

Nevertheless, we believe that the analysis strategy we have described 
can serve as a very useful starting point for the experimental search for 
disoriented chiral condensates. 
\\
\\
\begin{center}
{\Large ACKNOWLEDGMENTS}
\end{center}

This work was supported in part by the U.S. Department of Energy, the U.S.
National
Science Foundation, the Guggenheim Foundation, the Timken Foundation, 
 the Ohio Supercomputer Center, and the
Case Western Reserve University Provost's Fund.
\\
\\
\\
{\Large APPENDIX: DCC DISTRIBUTIONS}\\
\\
The distribution 
 \begin{displaymath}
 P[n;N]=\left( {{{2^{N}N! } \over {2^{n}n! }}} \right)^{2}{{(2n)! } \over
{(2N+1)! }}, \; \; \; \;  \; \; \; \; \; \; \; \;(A1)
\end{displaymath}
where $N$ and $n\leq N$ are nonnegative integers,
was discovered by Horn and Silver [15] in the context of coherent-state
production models. For
this reason we will refer to it as the {\em coherent} distribution. It was
later found that the coherent distribution was an appropriate final state
for a simple
 model of a zero isospin DCC [6].
 In both physical contexts, the distribution is relevant to the case of
 an even total number, $2N$, of pions and, necesarily, because of zero isospin,
to an even number, $2n$, of $\pi^{0}$'s. In the mathematical considerations
that follow, $n$ and $N$ are regarded as
arbitrary nonnegative integers.

In [6] it was shown that 
\begin{displaymath}
P[n;N]\rightarrow {1 \over {2\sqrt {f}}}{1 \over N},\; \; \; \; \; \;\; \;
\; \; \; \;\; \; \; \; \; \; \; \;\; \; \; \;\; \; \; \;(A2)
\end{displaymath} 
as $N,n\rightarrow \infty$, with $n/N\equiv f$ constant, in agreement with
the classical expectations
for a DCC [1-5, 7, 11]. Generally, a bivariate distribution can be
expressed as a continuous binomial 
distribution weighted over the infinite-sampling limit:
\begin{displaymath}
P[n;N]=\left( \begin{array}{c}
N\\
n
\end{array}
\right)\int_0^1 {{{df} \over {2\sqrt f}}}f^n(1-f)^{(N-n)}.\; \; \; \; \; \;
\; \;\; \; \; \;\; (A3)
\end{displaymath}
The representation (A3) can be established by rewriting (A1) in terms of
the beta function, 
\begin{displaymath}
             B(x,y)={{\Gamma (x)\Gamma (y)} \over {\Gamma (x+y)}}\; \; \;
\; \; \; \; \;\; \; \; \; \; \; \; \;\; \; \; \;(A4)
\end{displaymath}     
as
\begin{displaymath} 
P[n;N]={1 \over 2}\left( \begin{array}{c}
N\\
n
\end{array}\right)B(n+{1 \over 2},1+N-n),\; \; \; \; \; \; \; \;(A5)
\end{displaymath} 
which is found using the identity
\begin{displaymath}
\Gamma \left( {n+{1 \over 2}} \right)={{(2n)! } \over {2^{2n}n! }}\Gamma
\left( {{1 \over 2}} \right). \; \; \; \;\; \; \; \; \; \; \; \;(A6)
\end{displaymath}
The standard integral representation of the beta function [36] yields (A3).
The identity (A5) establishes the connection between the coherent-state
production model of 
Martinis, {\em et al.}, [9], for $I=0$ and the analysis of [6] and [15].
The continuous binomial distribution (A3) allows one to calculate all 
averages in the same explicit manner as for the binomial distribution for a
particular $f$ and then 
integrate the result over $f$ with the indicated weighting leading to exact
results for the 
various moments. The direct use of (A1) to calculate averages is awkward.

A problem arises with the original interpretation of distribution (A1) in
connection with a 
realistic dectector, or, equivalently, a sampling consisting of a finite
number of pions. 
The limited sampling of such a detector means that typically one sees only
a portion of the
particular group of the correlated pions that are thought to be the earmark
 of a DCC.  Within that sampling we need to find the distribution induced
by the DCC and with it
we can carry out a generating-function analysis. We show that the coherent
distribution is
self-similar in that the neutral/charged distribution of a finite number of
pions chosen from a sampling space distributed using the limit of (A1) for
$N\rightarrow\infty$, is given in fact by (A1), but now with $N$ and $n$
regarded as
the total number and the number of $\pi^{0}$'s, respectively, whether or
not they are even or odd.

In support of these remarks, let us consider the problem of the neutral/charged 
distribution of an
 arbitrary subset, even or odd, of a DCC corresponding to $2N$ pions that are 
distributed according to $P[n; N]$.  Suppose then that because of limited 
sampling we observe  $n_{t} \le 2N$ pions.  The joint probability 
distribution function for finding  $n_{0}$  neutral pions  and 
$n_{ch}=n_{t}-n_{0}$ 
charged pion's is then a product with $P[n,N]$ of the hypergeometric
distribution [35] of the
two relevant binomial samplings: 
\begin{displaymath}
Q[n_{0};n_{t};N]=\sum\limits_{n\ge {1 \over 2}n_{0}}^{N-{1 \over 2}n_{c}}
{\left( \begin{array}{c}
2n\\
n_{0}
\end{array}
\right)}\left( \begin{array}{c}
2(N-n)\\
n_{ch}
\end{array}
\right)\left[ {\left( \begin{array}{c}
2N\\
n_{t}
\end{array}
\right)} \right]^{-1}P[n; N],\; \; \; \; \; \; \; \;(A7)
\end{displaymath}                                                          
               
where realizing equality in either of the limits is possible only when these
 limits are even.  The nature of the summation limits in (A7) complicates a
 direct proof of the correct normalization, viz.,  
\begin{displaymath}  
\sum\limits_{n_{0}=0}^{n_{t}} Q[n_{0}; n_{t};N]=1,\; \; \; \; \; \; \; \;\;
\; \; \;(A8)
\end{displaymath}
however, (A8) has been verified numerically.

	Because of the limited sample one cannot regard $N$ in (A7) as known.
Therefore the
 case where all that is known is that $N\gg 1$  is of special interest. In
this case we find
using (A2), the Stirling approximation, and passing 
to the continuum limit of $f$, that
\begin{displaymath} 
Q[n_{0}; n_{t};N] \rightarrow P[n_{0}; n_{t}], \; \; \; \; \; \; \; \;\; \;
\; \;(A9)
\end{displaymath} where
\begin{displaymath}
P[n_{0}; n_{t}]=\left( \begin{array}{c}
                  n_{t}\\
                  n_{0}
\end{array}
\right)\int_{0}^{1} {{{df} \over
{2\sqrt{f}}}}f^{n_{0}}(1-f)^{(n_{t}-n_{0})},\; \; \; \; \; \; \; \;(A10)
\end{displaymath}

which has precisely the same form as (A3). Here, however, the
 respective functional parameters are the number of neutral and total pions 
sampled from the DCC, rather than half those numbers as they are for all of
 the pions of a full DCC. Thus the induced representation (A10) is a
quasi-coherent
 distribution that goes over to the classical DCC distribution (A2) in in
the infinite-$n_{t}$ limit, 
which in 
practice may not be too large, due to the accuracy of the Stirling
approximation for fairly small
 numbers.

  	The similar forms of Eqs. (A3) and (A10) shows that, in regard to an 
infinite sampling space, the coherent distribution generates a self-similar 
induced distribution.  In addition, the procedure used to arrive at (A10) 
indicates how one uses the continuum limit of the coherent distribution to
define
 a sampling of a finite number of pions from an infinite sampling space.  
This remark then also explains the use of the form (A3) when it is applied
to the full DCC: 
It represents a sampling algorithm carried out by means of neutral pairs of
pions to induce
 a DCC of a finite, even number of pions out of the infinite sample. 

	The distribution (A10) refers to a collection of pions that need not have
 net zero charge, the signal characteristic of  a full DCC, but yet makes no
 reference to the total charge.  For the sampling algorithm used to obtain 
(A10), the absolute magnitude of total charge will obviously be binomially 
distributed about zero if $n_{ch}$ is even, and unity if it is not; this 
extended form of (A10) should be used when the sign of the pions can be 
distinguished.  When they cannot, the means and variances have
interpretations that are
 different from a DCC.

Finally, let us weight $P[n_{0}; n_{t}]$ with
respect to a parent distribution $P[n_{t}]$. Then the relevant generating
function is 
\begin{displaymath}
 G_{DCC}(z_{ch},z_{0})=\sum\limits_{n_{t},n_{ch},n_{0}=0}^{\infty} 
{\delta_{n_{t},n_{ch}+n_{0}}}P[n_{0};
n_{t}]P[n_{t}]z_{ch}^{n_{ch}}z_{0}^{n_{0}}.\; \; \; \; \; \; \; \; \; \; \;
\;(A11)
\end{displaymath} 
Representation (A10) when combined with (A7) yields 
\begin{displaymath} 
G_{DCC}(z_{ch},z_{0})=\int_0^1 {{{df} \over {2\sqrt
{f}}}}G_{Bin}(z_{ch},z_{0};f),\; \; \; \; \; \; \;\; \; \; \; \;(A12) 
\end{displaymath}
which we interpret as the generating function of the factorial moments of
the numbers of charged and 
neutral pions sampled from a very large DCC sample space.

The distribution $p(f) = 1/(2\sqrt {f})$ has been associated with the 
decay of a DCC in the classical limit. Thus, the generating function (A12)
can be 
considered applicable to the situation in which the phase-space domain of
the particles 
resulting from the DCC is very much larger than the acceptance of the
detector. Then one can 
picture DCC production as corresponding to an event distribution for which
the neutral 
fraction $f$ is a random variable distributed according to $1/(2\sqrt
{f})$, a depiction 
reflected in (A12).
\\
\\
\\
\\
\section{REFERENCES}
\begin{enumerate}
\item A. A. Anselm, Phy. Lett. {\bf B217},169 (1989).
\item A. A. Anselm and M. G. Ryskin, Phy. Lett. {\bf B266}, 482 (1991).
\item J. D. Bjorken, SLAC-PUB-5545, Int. J. Mod. Phys. {\bf A7}, 4189 (1992).
\item J. D. Bjorken, Acta Phys. Pol. {\bf B23}, 561 (1992).
\item J-P. Blaizot and A. Krzywicki, Phys. Rev. D {\bf 46}, 246 (1992).
\item K. L. Kowalski and C. C. Taylor, CWRUTH-92-6, hep-ph/9211282.
\item K. Rajagopal and F. Wilczek Nucl. Phys. {\bf B399}, 395 (1993).
\item J. D. Bjorken, K. L. Kowalski and C. C. Taylor, SLAC-PUB-6109, Proc.
of Les Rencontres
de la Vall\'{e}e D' Aoste, La Thuile, 1993, ed. M. Greco, Editions
Frontier, p. 507 (1993).
\item M. Martinis, V. Mikata-Martinis, A. \u{S}vorc, and J. \u{C}rnugelj, Phys.
Rev. D {\bf 51}, 2482 (1995); hep-ph/9411329; hep-ph/9501210.
\item R. D. Amado {\em et al.}, Phys. Rev. Lett. {\bf 72}, 970 (1994).
\item B. M\"{u}ller, Rep. Prog. Phys. {\bf 58}, 611 (1995).
\item For a review, K. Rajagopal, in Quark-Gluon Plasma 2, ed. R. Hwa,
World Scientific, 1995,
 HUTP-95-A013.
\item I. V. Andreev, JETP Lett. {\bf 33},  67 (1981).
\item V. Karmanov and A. Kudrjavtsev, ITEP-88, 1983.
\item  D. Horn and R. Silver, Ann. Phys. (N.Y.) {\bf 66}, 509 (1971).
\item P. Carruthers and C. C. Shih, Int. J. of Mod. Phys. A {\bf 2}, 1447
(1987).
\item C. Geich-Gimbel, Int. J. of Mod. Phys. A {\bf 4}, 1527 (1989).
\item I. M. Dremin, Mod. Phys. Lett. {\bf A8}, 2747 (1993); Pis'ma Zh.
Eksp. Teor. Fiz. 
{\bf 59}, 561 (1994) [trans.
JETP Lett. {\bf 59}, 585 (1994)].
\item I. M. Dremin and R. C. Hwa, Phys. Rev. D {\bf 49}, 5805 (1994).
\item For a review, E. A. DeWolf, I. M. Dremin, W. Kittel, hep-ph/950325.
\item G. H. Thomas and B. R. Webber, Phys. Rev. D {\bf 9}, 3113 (1974).
\item L. Di\'{o}si, Nucl. Instrum. Methods {\bf 138}, 241 (1976).
\item L. Di\'{o}si, Nucl. Instrum. Methods {\bf 140}, 533 (1977).
\item L. Di\'{o}si and B. Luk\'{a}cs, Phys. Lett. {\bf B206}, 707 (1988).
\item R. E. Ansorge {\em et al.}, Z. Phys. {\bf C 43}, 75 (1989).
\item For descriptions and preliminary results of the 
MiniMax experiment (T-864) see: J. D.Bjorken {\em et al}, Fermilab Proposal
T-864, April 1993.
J. D. Bjorken, K.L. Kowalski, and C. C. Taylor, in Proc. of the 1993
Madison/Argonne
Workshop on Physics at Current  Accelerators  and the Supercollider, ed. J.
L. Hewett,
A. R. White, and D. Zeppenfeld,  p. 73, hep-ph/9309235. J. D. Bjorken,
SLAC-PUB-6430-1994.
C. C. Taylor, in Proc. Inter. Conf. on Elastic and Diffractive Scattering, ed.
H. M. Fried, K. Kang, and C-I Tan, World Scientific, 1994, p. 348.
C. C. Taylor, in Hot Hadronic Matter: Theory and Experiment, ed. J. Letessier
 {\em et al.},
Plenum Press, NY, 1995, p. 503.
W. L. Davis {\em et al}, in Proc. of the 24th Int. Cosmic Ray Conf.,
 Rome, September 1995. C. C. Taylor, Eialot, 1995. M.E. Convery {\em et al}, 
Bull. Am. Phys. Soc. {\bf 41}, 902 (1996). W. L. Davis {\em et al}, 
Bull. Am. Phys. Soc. {\bf 41}, 938 (1996). J. D. Bjorken, proceedings of the 
Nijmigen workshop, 1996.
C. C. Taylor, proceedings of the RHIC workshop, 1996. J. Streets, 
Bull. Am. Phys. Soc.  1996; Proc. of the DPF meeting, 1996.  
\item J. Pumplin, Phys. Rev. D {\bf 50}, 6811 (1994).
\item I. S. Gradshteyn and I. M. Ryzhik, {\em Table of Integrals, Series,
and Products}, 
Academic Press, Orlando, 1980, p. 20.
\item C. M. G. Lattes {\em et al.}, Phys. Rep. {\bf 65}, 151 (1980); 
L.T. Baradezi {\em et al.}, Nucl. Phys. {\bf B370}, 365 (1992); S-I
Hasegawa, ICR-151-87-5, 
unpublished.
\item G. Alverson {\em et al.}, Phys. Rev. D {\bf 48}, 5 (1993).
\item A. L. S. Angelis {\em et al.}, CERN/SPSLC/91-17,  May 1991;
CERN/SPSLC/95-35, May 1995, 
WA98 collaboration. B. Wyslouch, for the WA98 collaboration, talk at ICHEP,
July 25 1996. 
\item H. -U. Bengtsson and T. Sj\"{o}strand, Comp. Phys. Comm. {\bf 46}, 43
(1987).
\item GEANT-Detector and Simulation Tool, CERN, PM0062 (1993).
\item C. J. Liapis, Yale University thesis, May, 1995. CERN P238
Collaboration, J. Ellett 
{\em et al.}, to be published.
\item A. G. Frodesen, O. Skjeggestad, and H. T\o fte, {\em Probability and
Statistics in Particle Physics}, Universitetsforlaget, Bergen, 1979.
\item A.Erd\'{e}lyi, Higher Transcendental Functions, McGraw-Hill, New
York, 1953,
 Vol. 1, p. 10.

\end{enumerate}

\vfill\eject

{\bf Table 1.} Robust observables $r_{i,j}$ for generic events simulated by 
PYTHIA and pure DCC events simulated with 
the 'snowball' model. Comparisions with the $r_{i,j}$'s obtained with
binomially distributed pions 
and the $1/(2\sqrt{f})$ classical limit of DCC's.

\vskip1ex
\begin{tabular}{cccccc}
     &    & $\quad$PYTHIA$\quad$ & 
  $\quad\ $DCC$\ \ \quad$ & binomial & $1/(2\sqrt{f})$ \\
 $i$ & $j$ & $r_{i,j} \pm \sigma_{r_{i,j}}$ &  
$r_{i,j} \pm \sigma_{r_{i,j}}$ &
$r_{i,j}$ & $r_{i,j}$ \\
\hline
   1 & \ 1 \ &  $\quad$1.00 $\pm$ 0.02$\quad$ &  $\quad$0.57 $\pm$
0.01$\quad$ & 1.00 & 0.50 \\
   2 & \ 1 \ &  $\quad$1.00 $\pm$ 0.05$\quad$ &  $\quad$0.43 $\pm$ 0.03$\quad$ & 1.00 & 0.33 \\
   3 & \ 1 \ &  $\quad$1.04 $\pm$ 0.13$\quad$ &  $\quad$0.38 $\pm$
0.05$\quad$ & 1.00 & 0.25 \\

   0 & \ 2 \ &  $\quad$1.36 $\pm$ 0.04$\quad$ &  $\quad$1.55 $\pm$
0.06$\quad$ & 1.36  & 1.80 \\
   1 & \ 2 \ &  $\quad$1.36 $\pm$ 0.10$\quad$ &  $\quad$0.66 $\pm$
0.06$\quad$ & 1.30  & 0.62 \\
   2 & \ 2 \ &  $\quad$1.47 $\pm$ 0.26$\quad$ &  $\quad$0.44 $\pm$
0.09$\quad$ & 1.25  & 0.31 \\

   0 & \ 3 \ &  $\quad$2.13 $\pm$ 0.25$\quad$ &  $\quad$2.98 $\pm$
0.39$\quad$ & 1.89  & 3.54 \\
   1 & \ 3 \ &  $\quad$2.03 $\pm$ 0.43$\quad$ &  $\quad$1.19 $\pm$
0.31$\quad$ & 1.74  & 0.90 \\
 
   0 & \ 4 \ &  $\quad$3.06 $\pm$ 0.94$\quad$ &  $\quad$6.82 $\pm$
2.18$\quad$ & 2.70  & 7.34 \\
\end{tabular}

\vfill\eject

{\bf Table 2.} The effect on the $r_{i,1}$ of an admixture
 of DCC and generic (PYTHIA) events. DCC domains from the DCC-generator/GEANT 
are added to 
various fractions of random PYTHIA/GEANT events. The first column 
represents the fraction of events in which a DCC 
is overlaying a generic event.

\vskip1ex
\begin{tabular}{ccccc}
 DCC fraction & $\quad r_{1,1} \pm \sigma_{r_{1,1}}\quad$ &  
$\quad r_{2,1} \pm \sigma_{r_{2,1}}\quad$ &
$\quad r_{3,1} \pm \sigma_{r_{3,1}}\quad$ & $\quad$events$\quad$ \\
\hline
0.00 &  1.01 $\pm$ 0.02 &  1.02 $\pm$ 0.05 &  1.09 $\pm$ 0.14 &  51741 \\
0.02 &  1.00 $\pm$ 0.02 &  1.00 $\pm$ 0.05 &  1.01 $\pm$ 0.15 &  51741 \\
0.05 &  0.97 $\pm$ 0.02 &  0.93 $\pm$ 0.05 &  0.95 $\pm$ 0.10 &  51741 \\
0.10 &  0.95 $\pm$ 0.02 &  0.89 $\pm$ 0.04 &  0.89 $\pm$ 0.08 &  51741 \\
0.20 &  0.93 $\pm$ 0.02 &  0.83 $\pm$ 0.04 &  0.77 $\pm$ 0.07 &  51741 \\
0.50 &  0.84 $\pm$ 0.01 &  0.71 $\pm$ 0.03 &  0.68 $\pm$ 0.06 &  40000 \\
1.00 &  0.74 $\pm$ 0.01 &  0.60 $\pm$ 0.03 &  0.55 $\pm$ 0.06 &  20000 \\
\end{tabular}

\end{document}